\documentclass[conference,compsoc]{IEEEtran}
\usepackage{graphicx} 
\usepackage{balance}
\usepackage{booktabs}
\usepackage{subcaption}
\usepackage{multirow}
\usepackage{amsmath}
\usepackage{url}
\usepackage{hyperref}
\hypersetup{
    hidelinks=true
}
\usepackage{cleveref}
\usepackage{orcidlink}
\usepackage{tikz}
\usetikzlibrary{arrows.meta,positioning,fit,calc}

\ifCLASSOPTIONcompsoc
  \usepackage[nocompress]{cite}
\else
  \usepackage{cite}
\fi
\ifCLASSINFOpdf
\else
\fi
\newcommand{\cherid}{\mbox{CHERI-\textbf{\textit{D}}}}

\begin{document}
%
\title{\cherid{}: Secure and efficient inline object ID for\\ CHERI temporal memory safety }

\makeatletter
\newcommand{\linebreakand}{%
  \end{@IEEEauthorhalign}
  \hfill\mbox{}\par
  \mbox{}\hfill\begin{@IEEEauthorhalign}
}
\makeatother

\author{\IEEEauthorblockN{Yuecheng Wang
\orcidlink{0009-0009-8075-7006}}
\IEEEauthorblockA{University of Cambridge\\
United Kingdom\\
yuecheng.wang@cl.cam.ac.uk}
\and
\IEEEauthorblockN{Jonathan Woodruff
\orcidlink{0000-0003-3971-2681}}
\IEEEauthorblockA{University of Cambridge\\
United Kingdom\\
jonathan.woodruff@cl.cam.ac.uk}
\and
\IEEEauthorblockN{Alfredo Mazzinghi
\orcidlink{0009-0009-2941-4729}}
\IEEEauthorblockA{University of Cambridge\\
United Kingdom\\
alfredo.mazzinghi@cl.cam.ac.uk}
\linebreakand
\IEEEauthorblockN{Peter Rugg
\orcidlink{0009-0000-2976-0474}}
\IEEEauthorblockA{University of Cambridge\\
United Kingdom\\
peter.rugg@cl.cam.ac.uk}
\and
\IEEEauthorblockN{Samuel W. Stark
\orcidlink{0000-0002-7268-9471}}
\IEEEauthorblockA{University of Cambridge\\
United Kingdom\\
samuel.stark@cl.cam.ac.uk}
\and
\IEEEauthorblockN{Alexandre Joannou
\orcidlink{0000-0002-3161-2638}}
\IEEEauthorblockA{University of Cambridge\\
United Kingdom\\
alexandre.joannou@cl.cam.ac.uk}
\linebreakand
\IEEEauthorblockN{Robert N. M. Watson
\orcidlink{0000-0001-8139-8783}}
\IEEEauthorblockA{University of Cambridge\\
United Kingdom\\
robert.watson@cl.cam.ac.uk}
\and
\IEEEauthorblockN{Simon W. Moore
\orcidlink{0000-0002-2806-495X}}
\IEEEauthorblockA{University of Cambridge\\
United Kingdom\\
simon.moore@cl.cam.ac.uk}}


%


    \maketitle

\begin{abstract}

We propose \cherid{}, an architectural extension to CHERI that supports efficient temporal memory safety.
Efficient memory safety is an increasing priority for programming languages, operating systems, and hardware designs, and CHERI is a leading hardware/software system that provides native spatial safety and a foundation for temporal memory safety~\cite{cheriv9}. 
Due to CHERI lack of intrinsic architectural support for temporal memory safety, the state-of-the-art CHERI temporal safety solution, Cornucopia Reloaded, is a software-based solution that provides use-after-reallocation (UAR) protections instead of the stronger use-after-free (UAF) mitigation, and suffers performance overhead due to delayed reallocation and revocation. 
\cherid{} associates object identification (ID) metadata with capability pointers to provide temporal integrity of allocations.
\cherid{} leverages CHERI's spatial protection to safely co-locate trusted temporal metadata with application objects, relaxing the conventional requirement for a separate shadow metadata structure and allowing object IDs to be stored inline, potentially within otherwise-unused allocation fragmentation.
Evaluated in simulation and in hardware, \cherid{} significantly reduces the revocation overhead of Cornucopia Reloaded while allowing it to support strict use-after-free mitigation.

\end{abstract}


%
\IEEEpeerreviewmaketitle

\section{Introduction}
CHERI is a promising technology to mitigate memory safety vulnerabilities~\cite{cheri_security_analysis}.
CHERI provides spatial safety by constraining the memory access range of pointers within bounds that cannot be forged or increased.
However, since pointers are allowed to be freely copied and distributed, it is challenging to enforce temporal safety by revoking memory access when the lifetime of an allocation is over.

The state-of-the-art CHERI temporal safety solution, Cornucopia Reloaded~\cite{Cornucopia_reloaded}, employs a quarantine-and-sweep-based approach that delays reallocation until a sweep of the address space has revoked dangling heap capabilities.
Similarly to other software-based approaches~\cite{ainsworthMarkUsDropinUseafterfree2020, erdosMineSweeperCleanSweep2022}, it mitigates \emph{use after reallocation} (UAR) rather than providing a stronger guarantee of \emph{use after free} (UAF) protection.

Due to the lack of architectural support in CHERI to represent capability lifetime, Cornucopia Reloaded must buffer freed memory in quarantine rather than immediately returning it to the allocator.
Unfortunately, memory quarantining can significantly affect allocator optimization, memory footprint, and cache efficiency~\cite{wangPoisonCapEfficientHierarchical2026, wesleyfilardoCornucopiaTemporalSafety2020}.
When added to the cost of revocation sweeps, overall system performance and resource usage are impacted.

We propose a novel architectural extension to CHERI, \cherid{} (pronounced, ``cherry-dee"), that supports a stronger temporal safety guarantee at much greater efficiency.
\cherid{} introduces architectural support for fine-grained tracking of pointer/memory lifetime ({\it i.e.,} generation) to provide temporal guarantees.
\cherid{} assigns an architecturally protected lifetime identifier to each allocated object.
As a result, a single memory location may be issued under multiple allocation lifetimes, with the previous lifetime identifier invalidated when the object is freed.
This mechanism enables immediate and safe memory reuse while preserving strong temporal safety guarantees.
In contrast to previous approaches that depend on delayed reuse using quarantine buffers and costly revocation mechanisms~\cite{ainsworthMarkUsDropinUseafterfree2020,erdosMineSweeperCleanSweep2022, wesleyfilardoCornucopiaTemporalSafety2020, Cornucopia_reloaded, CHERIoT}, our design allows reclaimed memory to be promptly returned to the allocator without exposing newly allocated objects to dangling pointers.

More broadly, \cherid{} demonstrates how capability architectures change the design space for trusted metadata.
Conventional designs have to isolate trusted metadata from application memory to prevent corruption.
CHERI's fine-grained capability bounds relax this requirement: metadata can be placed adjacent to an object while remaining outside the bounds accessible to the application.
\cherid{} exploits this property to protect object-local lifetime metadata, allowing CHERI's spatial-safety guarantees to directly support temporal safety.

This property enables \cherid{} to avoid the fixed-granularity metadata organization used by MTE-like approaches~\cite{ARM_MTE,ApplicationDataIntegrity}.
Rather than replicating metadata across memory granules, \cherid{} associates a single ID with each allocation and can place it inline with allocation data.
In many cases, \cherid{} IDs incur no memory overhead when placed by the allocator in unused internal fragmentation and padding.
\cherid{}'s more efficient metadata representation enables a larger per-object lifetime space than MTE; our prototype supports 254 IDs for immediate reuse before falling back to quarantine and sweeping.

We implemented a prototype of \cherid{} that associates an 8-bit ID with each allocation slot.
\cherid{} encodes both the capability ID and the memory location of the current slot ID in protected capability metadata; this metadata allows hardware to load and verify the ID when the capability is used.
Using this system, memory regions can be reclaimed immediately after deallocation and safely reallocated through 254 generations, thus improving memory reuse efficiency while reducing the growth rate of the quarantine buffer.
Our evaluation demonstrates that \cherid{} substantially reduces revocation overhead while preserving stronger temporal memory safety guarantees.

In this work, we make the following contributions.
\begin{itemize}
    \item An analysis of memory allocation patterns from a selection of real-world applications for efficient architectural design choices. 
    \item An architectural extension, \cherid{}, that exploits CHERI spatial protection to safely co-locate trusted per-object lifetime metadata with application memory, providing use-after-free protection while reducing quarantine and revocation overhead.
    \item Implementations of \cherid{} prototypes in the CHERI-RISC-V QEMU full system emulator~\cite{qemu} and in the CHERI-Toooba CPU, a superscalar, out-of-order FPGA soft-core~\cite{ruggSuiteProcessorsExplore2024}. 
    \item \cherid{} software support in the CHERI-enabled LLVM/clang toolchain~\cite{llvm}; the CHERI-enabled FreeBSD operating system,  CheriBSD; and in the Malloc Revocation Shim (MRS) wrapper~\cite{cheribsd} of the CheriBSD system allocator jemalloc~\cite{Jemalloc}.
    \item Demonstrating that \cherid{} successfully detects and mitigates use-after-free and double free in a complete set of 2422 NIST Juliet test cases ~\cite{Juliet}, and showing the impact of \cherid{} on revocation performance using SPEC CPU2006 INT, SQLite and gRPC for evaluation. ~\cite{henningSPECCPU2006Benchmark2006, sqlite, GrpcGrpcV1542}. 

\end{itemize}

\section{Background}

\subsection{Cornucopia Reloaded}

The state-of-the-art CHERI temporal safety system is called ``Cornucopia Reloaded'' (or just \emph{Reloaded})~\cite{Cornucopia_reloaded}.
Reloaded scans the address space to revoke capabilities pointing to freed memory.
Due to a lack of architecture support for temporal safety, Reloaded is primarily a software system, though it relies on new page table bits to allow concurrent revocation scans.
For Reloaded, the memory allocator must quarantine freed allocations, painting them in a software-managed shadow bitmap with one bit per 16-byte word in memory.
When quarantined memory reaches a percentage threshold of the heap (typically 25\%), the allocator triggers a revocation sweep by the Reloaded system service.
The size of the quarantine buffer can rapidly grow in allocation-intensive benchmarks and applications~\cite{471omnetppSPECCPU2006,sqlite}, leading to frequent revocation runs.
Although Cornucopia Reloaded runs revocation concurrently on a separate CPU core to the application core to avoid revocation overhead, delayed memory reuse and the background scan increase cache pollution and DRAM bandwidth, resulting in performance overhead.

Reloaded's strategy of memory quarantining and sweeping is a common strategy to support temporal safety for application-class systems~\cite{ainsworthMarkUsDropinUseafterfree2020, erdosMineSweeperCleanSweep2022, CHERIoT}.
This strategy is favored because it has minimal impact on common-case memory access.
However, these solutions primarily provide use-after-reallocation mitigation rather than stronger use-after-free guarantees.
That is, memory accesses to quarantined regions are permitted until revocation finishes, which can still leave a temporal window during which dangling pointers remain exploitable.
As argued in previous work, closing the window is necessary to fully eliminate use-after-free vulnerabilities~\cite{wangPoisonCapEfficientHierarchical2026}.

\subsection{Memory quarantining}

The cost of delayed reallocation ({\it i.e.,} quarantining) in real-world applications has not yet been fully explored and understood.
Delayed reallocation increases the memory footprints of applications, which increases system memory overhead and cache pressure.
Delayed reallocation disrupts the fast-path optimizations of modern allocators by breaking their assumptions about the immediate reuse of recently freed allocations~\cite{Jemalloc,snmalloc,hunterMallocEfficiencyFleet}, which is central to thread-local caching mechanism such as jemalloc's tcache and similar LIFO schemes.
In terms of jemalloc, batching the return of freed memory thwarts thread-local bins, causing tcache misses and forcing it to fall back to slower arena-level refills and synchronized heap structures.
This increases lock contention, cache, and TLB pressure, and reduces slab and CPU-cache locality.

\subsection{Memory metadata}

Temporal safety can be enforced by using metadata to associate pointers and memory.
Architectures such as SPARC ADI and ARM MTE exemplify this approach by associating metadata ({\it i.e.,} \emph{tags}) with memory locations and validating accesses against matching metadata in pointers at runtime~\cite{ARM_MTE, ApplicationDataIntegrity}. 
Ideally, a memory access is considered valid only if the pointer tag matches the current memory tag; that is, the access occurs within the allocation's lifetime.

However, associating metadata with memory introduces both security and performance challenges.
Metadata integrity must be preserved, and existing approaches commonly rely on shadow memory to store metadata, increasing memory overhead and complicating memory management~\cite{ARM_MTE, ApplicationDataIntegrity}.
Furthermore, existing designs typically support metadata at a fixed granularity ({e.g.} 4 bits per 16 bytes), such that metadata overhead scales linearly with allocation size.
Runtime metadata validation may also introduce performance and hardware area overhead, and additional metadata lookup structures and caches are required to support efficient checking.
Moreover, limited tag storage and lack of pointer integrity lead to only probabilistic protection in existing tagging schemes, which lack strong security guarantees~\cite{ARM_MTE, ApplicationDataIntegrity}. 

\begin{table*}[t]
\centering
\caption{
Heap allocation size distribution (Dist.) and fraction of allocations
with internal fragmentation (IF.) under jemalloc. For SPEC CPU2006 INT,
slash-separated values report train/ref inputs, respectively; SQLite
reports a single speedtest1 measurement. Percentages are rounded.
}
\label{table:allocation_distribution}

\setlength{\tabcolsep}{4pt}

\begin{tabular}{
l
cc
cc
cc
cc
cc
c
}
\toprule

\multirow{2}{*}{Benchmark(train/ref)}
& \multicolumn{2}{c}{\textless 128 B (\%)}
& \multicolumn{2}{c}{\textless 512 B (\%)}
& \multicolumn{2}{c}{\textless 4 KiB (\%)}
& \multicolumn{2}{c}{\textless 4 MiB (\%)}
& \multicolumn{2}{c}{\textgreater 4 MiB (\%)}
& \multirow{2}{*}{Allocations}
\\

& Dist. & IF.
& Dist. & IF.
& Dist. & IF.
& Dist. & IF.
& Dist. & IF.
&
\\

\midrule

bzip2
& 0/0 & 0/0
& 0/0 & 0/0
& 0/0 & 0/0
& 90/90 & 89/89
& 10/10 & 0/0
& 30/30
\\

gobmk
& 44.5/38 & 35.3/36
& 0.01/0.01 & 100/100
& 0.06/0.06 & 100/100
& 55.4/61.9 & 99.9/99
& 0.01/0.01 & 66.7/67
& 51067/133215
\\

hmmer
& 8.6/7 & 93.6/94
& 49.1/40 & 98.1/98
& 42.2/53 & 99.1/99
& 0.04/0.01 & 93.4/93
& 0/0 & 0/0
& 170127/1.0M
\\

sjeng
& 0/0 & 0/0
& 0/0 & 0/0
& 0/0 & 0/0
& 33.3/33.3 & 0/0
& 66.7/66.7 & 100/100
& 6/6
\\

libquantum
& 35.8/34 & 5.1/4
& 3.7/3 & 0/0
& 6/4 & 0/0
& 55.1/37 & 31.7/51
& 0/23 & 0/15
& 109/150
\\

h264ref
& 19/19 & 83/83
& 38/32 & 97/97
& 36/36 & 100/100
& 13/13 & 100/100
& 0/0 & 0/0
& 38271/38271
\\

omnetpp
& 22/18.8 & 87/7.5
& 78/81.2 & 76.2/98.8
& 0.01/0.01 & 96/99.7
& 0.01/0.01 & 12.6/86.84
& 0/0 & 0/0
& 130M/267M
\\

xalancbmk
& 95/68.0 & 74.9/51
& 0.9/8.4 & 88.1/95.2
& 1.1/21.2 & 39.1/98.9
& 3.1/1.6 & 99.7/99.7
& 0/0 & 0/0
&  1.1M/135M
\\

sqlite (speedtest1)
& 27.9 & 0
& 5.7 & 53.6
& 0.6 & 70
& 65.8 & 52.8
& 0 & 0
& 221129
\\

\bottomrule
\end{tabular}
\end{table*}

\subsection{CHERI with memory versioning}

To mitigate the performance overhead associated with memory quarantining and frequent sweeps, Microsoft proposed combining CHERI with MTE-based memory versioning~\cite{DiscussionMSRsCHERI+MTE}.
Traditional ARM MTE provides only probabilistic temporal safety detection because tag aliasing can occur after approximately 16 memory reuses.
In this approach, MTE tags are repurposed as version identifiers to reduce the security gap for use-after-free (UAF) and use-after-reallocation (UAR) vulnerabilities while reducing revocation frequency.
With a 4-bit MTE tag space, Cornucopia Reloaded can defer revocation until version exhaustion occurs.

However, the ARM-MTE style of memory versioning has several limitations that prevent it from being scalable for large and allocation-intensive workloads.
First, memory versioning fundamentally trades off metadata overhead against version space size.
Existing ARM MTE associates a 4-bit tag with every 16-byte memory granule~\cite{ARM_MTE}, incurring approximately 3.12\% memory overhead while allowing only 16 unique versions before tag aliasing occurs.
This version space scales badly for several allocation-intensive CHERI workloads, some of which trigger more than 2000 revocations, as shown in Table~\ref{tab:revocation-quarantine}.
Increasing the width of the tag can further reduce the revocation frequency, but this incurs additional memory and hardware overhead to store and track the tags. 
Second, previous memory versioning designs scale poorly for large allocations because version metadata is stored at a fixed granularity~\cite{ARM_MTE, DiscussionMSRsCHERI+MTE}.
Incrementing the version of an allocation requires updating the tags of all covered memory granules, with a cost proportional to allocation size.

\subsection{Dynamic memory allocation}
\subsubsection{Application allocation patterns}
Applications exhibit distinct dynamic memory allocation behaviors depending on their workload characteristics.
As shown in Table~\ref{table:allocation_distribution} and~\ref{tab:revocation-quarantine}, allocation-intensive applications such as \textit{xalancbmk} and \textit{omnetpp} trigger a higher number of revocations compared to less allocation-intensive workloads.
This behavior arises because frequent allocation and deallocation activity causes the Cornucopia quarantine buffer to fill rapidly, triggering a large number of memory sweeps for revocation.
Furthermore, these workloads exhibit a high proportion of small-sized allocations.  

Such allocation patterns are commonly observed in server-client workloads such as gRPC~\cite{GrpcGrpcV1542}, and network-oriented applications, where execution is dominated by frequent, small allocations and deallocations.
For example, BSD networking stacks designed \textit{mbuf} allocation mechanisms based on the observation that network traffic exhibits a bimodal packet-size distribution; around 99\% of all TCP messages and 86\% of UDP messages are below 200 bytes~\cite{kayImportanceNondataTouching1993}.

Jemalloc follows a comparable design rationale by optimizing primarily for small-object allocations under the assumption that most applications allocate objects smaller than 512 bytes, as described in its original design~\cite{Jemalloc}.
Consequently, jemalloc distinguishes between small and large allocations, optimizing small allocations using slab-based bins organized within arenas, while large allocations are handled separately using dedicated allocation mechanisms.  

Motivated by the prevalence of small allocations in allocation-intensive workloads, this work optimizes for small memory allocations, in particular, for allocations that fit within a 4 KiB page.

\subsubsection{Alignment-driven memory layout: size classes and compiler padding}
\label{subsubsection:Alignment-driven memory layout}
In this work, we use internal fragmentation padding within allocation slots to store trusted metadata for temporal safety to avoid memory overhead. 

Memory alignment requirements are enforced by application binary interfaces (ABIs), compiler implementations, and hardware architectures, which require that objects be placed at memory addresses that meet specific alignment constraints for correct and efficient memory access.
Consequently, heap allocators return memory blocks that satisfy alignment requirements, typically defined by architecture-specific constraints.
This is commonly achieved by rounding the allocation sizes upward to the nearest alignment boundary.

Additionally, modern allocators employ size classes to improve allocation performance, predictability, and scalability by grouping allocation requests into fixed buckets, allowing for near constant-time allocation through per-class free lists or slab structures rather than exact-size matching ~\cite{wilsonDynamicStorageAllocation1995, Jemalloc, snmalloc, hunterMallocEfficiencyFleet}.
This approach improves throughput, cache locality, and concurrency but introduces internal fragmentation, since allocation requests are rounded up to the nearest size class, leaving unused spaces within allocated blocks.

In practice, the degree of internal fragmentation depends not only on the allocator design but also on the application memory behavior.
Many C and C++ applications dynamically construct irregular, pointer-rich, and input-dependent data structures whose allocation sizes vary substantially.
This behavior is observed in general-purpose workloads and benchmark suites, including SPEC CPU applications such as \textit{omnetpp} and \textit{xalancbmk}~\cite{471omnetppSPECCPU2006}, where various allocation sizes frequently interact with allocator size-class boundaries and expose allocator-induced rounding overhead, as shown in Table~\ref{table:allocation_distribution}.
In contrast, systems such as SQLite~\cite{sqlite} and gRPC~\cite{GrpcGrpcV1542} often perform application-level memory management through structured buffers, caches, or protocol-defined objects, constraining allocation to a smaller set of recurring sizes.
Although such designs may reduce allocator-induced fragmentation by aligning more consistently with allocator size classes, they may still introduce unused spaces internally through their own coarse-grained memory organization and object-level rounding. 

Consequently, internal fragmentation is not an allocator property, but an effect of allocator size classes and application-level memory structure.
More fundamentally, classical results in dynamic allocation theory show that perfect space utilization cannot be guaranteed under arbitrary allocation request sequences~\cite{robsonBoundsFunctionsConcerning1974}, implying that some degree of memory waste is unavoidable in general-purpose dynamic allocation systems. 

\subsubsection{Quarantine impact on allocator optimizations}
Modern allocators are generally designed to promote temporal locality; recently freed memory is prioritized to be reallocated in the near future.
For example, jemalloc aggressively reuses recently freed objects through its thread cache (tcache)~\cite{Jemalloc}.
Similar optimizations also exist in tcmalloc and other allocators~\cite{hunterMallocEfficiencyFleet}. 
However, memory quarantining delays reallocation and disrupts this assumption by preventing freed memory from being immediately returned to thread-local caches or allocation bins.
Instead, quarantine buffers hold freed objects until they can be safely reclaimed, increasing the likelihood that allocators fall back to slower allocation paths.
Even snmalloc, which is less dependent on this specific reuse assumption due to its design around per-core allocation and message-passing allocation, also experiences performance overhead under quarantining~\cite{snmalloc, wesleyfilardoCornucopiaTemporalSafety2020}.
In general, memory quarantining not only increases memory footprint but also disturbs allocation patterns, further increasing performance overhead.

\section{System and threat model}
\label{section:threat_model}
\cherid{} strengthens temporal safety relative to Cornucopia Reloaded by protecting against both use-after-free (UAF) and use-after-reallocation (UAR).
Cornucopia Reloaded prevents UAR by quarantining freed memory until dangling capabilities are revoked by a memory sweep, but does not prevent accesses to an object while it remains in quarantine.
\cherid{} instead invalidates dangling capabilities for access immediately on free by updating the object's memory ID, preventing UAF accesses before reallocation. Subsequent reallocation assigns a new ID, preventing stale capabilities from accessing the reallocated object, therefore also preventing UAR.
\subsection*{Trust model and software topology}
\cherid{} retains the software topology and trust model of Cornucopia Reloaded: the Malloc revocation shim (MRS)~\cite{Cornucopia_reloaded} and allocator remain user-space components, with no kernel privilege required for allocation.
MRS retains capabilities covering the underlying
allocator-managed allocations and returns IDed capabilities tightly bounded to the user-requested object sizes. 
ID metadata is accessed by MRS using these underlying capabilities; ordinary application capabilities cannot access this metadata because it lies outside their bounds. Thus, \cherid{} introduces neither privileged allocator operations nor additional kernel-resident allocator components.

Like prior work~\cite{CHERIvoke,CHERIoT,
wesleyfilardoCornucopiaTemporalSafety2020}, \cherid{} also focuses on user-space heap temporal safety. Extending these guarantees to kernel and stack memory is left for future work.

\section{\cherid{} overview}
\cherid{} extends CHERI to track allocation lifetime ({\it i.e.,} version or generation) in both capability pointers and memory.
We augment CHERI capabilities with an ID field representing the lifetime of memory references, while each memory slot in the allocator maintains a corresponding memory-resident ID of its current lifetime state.
Upon capability dereference, the ID stored in the capability is compared against the allocation's ID.
A mismatch indicates an out-of-lifetime access ({\it i.e.,} use-after-free violation) which is detected and mitigated by \cherid{}. 

\cherid{} stores the allocation ID within allocator-managed data.
Storing the ID inline with allocation data allows for efficient ID management in software and also enables efficient hardware lookup and checking.

\subsection{Design rationale}
\label{sec:design-rationale}
A key design goal of \cherid{} is to preserve CHERI's decentralized protection model while supporting strong temporal safety. Rather than introducing a centralized metadata table, \cherid{} associates temporal-safety metadata directly with individual allocations and places it close to the protected object. This allows metadata management to scale with allocation and memory capacity, without requiring a global structure shared across cores.

Co-locating IDs with object data also provides spatial locality.
\cherid{} therefore places an ID either within the allocation's cache line or within the same page as the object. 
Besides improving data-to-metadata locality, same-page placement allows the ID's physical address to be derived from the data address translation, avoiding an additional TLB translation for ID validation.
This decentralized organization may also reduce the shared-state
and synchronization bottlenecks associated with centralized metadata structures in future multicore and NUMA implementations.
\cherid{} also avoids MTE-style fixed-granularity, replicated metadata, but rather associates a single ID with each allocation object.
This allows richer metadata per object, as metadata storage does not grow proportionally with object size.
\cherid{}-style metadata also improves runtime, as freeing or reusing a large object requires updating only its ID rather than ``painting" metadata covering each memory granule.

Our allocation-trace analysis further shows that allocation-intensive workloads that suffer high revocation overhead are commonly dominated by small allocations, as shown in Table~\ref{table:allocation_distribution}. We therefore optimize the current \cherid{} design for objects up to 4 KiB, covering the common performance-critical case while keeping ID encoding and lookup simple. Support for larger allocations represents a different point in this design space and is discussed in future work.

\subsection{\cherid{} architecture extension}
\label{sec:architecture-extension}
\cherid{} maintains an ID for each memory allocation within capability metadata and also encodes the offset of the ID relative to the allocation.
Storing these in capability metadata allows for efficient ID management and checking. 

The \cherid{} architectural extension includes the following:
\begin{itemize}
    \item An 8-bit capability ID field represents the lifetime that a capability is authorized to access, as shown in Figure~\ref{fig:inlineid_mode} and Figure~\ref{fig:inpage_id_mode}. 
    \item A 1-bit capability ID mode (IDMODE) field differentiates capabilities using the inline ID storage scheme (\ref{subsection:inline_id}) from the in-page scheme (\ref{subsection:inpage_id}).
    \item A 6-bit capability ID location (IDLOC) field encodes the memory ID location according to the scheme selected by IDMODE. 
    \item Two new instructions, \emph{cgetcapID} and \emph{csetcapId}, read and set the \emph{ID} from and and in a capability.
    \item Two new instructions, \emph{csetIDloc} and \emph{cgetIDloc} set and read the a capability's \emph{IDLOC} field. The \emph{IDLOC} field will in-turn be used by the hardware on every future dereference of the capability to determine where the ID byte for this capability is stored. \emph{csetIDloc} therefore requires a second capability authorizing access to the memory ID location to be used, as described in Section~\ref{subsection:arch_id_protection}.
    \end{itemize}

These field widths represent the design point evaluated in our prototype rather than fundamental requirements of \cherid{}. We use an 8-bit \emph{ID} because one byte is the smallest naturally addressable metadata unit.
A  full byte provides 254 usable non-zero allocation generations; ID 255 is reserved for exhaustion/quarantine. 
\emph{IDLOC} is 6 bits firstly to allow full flexibility for placing the ID byte within a 64-byte line, and secondly to allow a distinct ID for each of 64 cache lines in a 4KiB page. 
\emph{IDMODE} is one bit because the current prototype implements two ID-placement modes: inline and in-page. This width of this field could potentially be extended to support more ID-placement modes. 
We encourage further quantitative evaluation of these parameters before committing them into production architecture.

At CPU reset, the \emph{root capabilities} are initialized with  capability ID, IDMODE and IDLOC set to zero.
This enables privileged code to delegate the authority to manage the \cherid{} mechanism (\ref{subsection:arch_id_protection}) and is consistent with the use of \textit{RESERVED-ZERO} bits of the capability encoding.
    
\subsection{\cherid{} protection model}
\label{subsection:arch_id_protection}

The CHERI protection model enforces monotonicity properties through the guarded manipulation of capabilities.
Similarly, \cherid{} preserves those properties when authorizing access to the memory IDs and manipulation of the \emph{ID}, \emph{IDLOC} and \emph{IDMODE} fields.

CHERI tightly bounds a capability to the user-requested allocation.
However, \cherid{} places the memory ID outside of the allocation bounds to protect it from modification by the program.
Therefore, the allocator (or another trusted component) must maintain an independent capability that authorizes access to the memory ID location so that the ID can be updated upon allocation and deallocation.

In addition, the modification of \emph{IDMODE} and \emph{IDLOC} fields must be subject to authorization checks.
Setting IDLOC authorizes reads of the ID byte in memory which likely lie outside the primary bounds of the capability returned to the application. Therefore, \texttt{csetIDloc} takes two capability operands: the capability whose IDLOC is to be updated, and a capability with read permission to the new ID location.
The \texttt{csetIDloc} instruction verifies that the memory-ID address represented by the new
\emph{IDLOC} lies within the bounds of the authority capability and that it permits reads. 
Indeed, both input capabilities must be valid, unsealed and have read permission, and the capability whose \emph{IDLOC} is modified must have a zero capability ID; any violation raises a capability exception. 

The \emph{ID} field seals the
\emph{ID}, \emph{IDMODE}, and \emph{IDLOC} fields: when the
capability ID is non-zero, attempts to modify these fields invalidate the capability. 
These constraints prevent application capabilities from modifying or redirecting the ID validation mechanism after it has been established by the allocator.

As a result, while \cherid{} capabilities unusually grant access to non-contiguous memory, this mechanism cannot escape the capability model, as \texttt{csetIDloc} ensures that read privilege for the ID location was available at the time \emph{IDLOC} was written, and \emph{ID} will seal these fields from further unexpected modification.

When adding these features to the architecture, extreme care needs to be taken to ensure that access to the ID bytes does not allow bypassing the existing capability model. We believe this is possible, e.g. by authorizing setting the ID byte with a capability that covers that byte.
However, we leave as further work ensuring that the capabilities are treated correctly in all corners of the architecture.
We recommend formal proof of the architecture that CHERI monotonicity properties (suitably tweaked) are preserved, e.g. as proved by \cite{bauereiss2022verified}.

To satisfy these constraints, the allocator must re-derive a capability for each reallocation in order to freely update the ID value.
This is not considered a limitation because most allocators already perform this re-derivation operation on the free path.
Inspecting the \emph{ID}, \emph{IDMODE}, and \emph{IDLOC} capability fields does not require additional authority; accesses to memory-resident IDs remain subject to the authorization requirements described above.

\subsection{ID Location}

\cherid{} decouples ID management from fixed memory granularities.
Unlike prior approaches that associate metadata with individual memory words or fixed-size memory granules~\cite{ARM_MTE, ApplicationDataIntegrity, Cornucopia_reloaded}, \cherid{} associates a single ID with each allocation.
Consequently, only one ID byte is required per object, and ID updates can be performed in constant time.
In comparison, granule-based schemes must update the metadata of all granules belonging to an object, causing the update overhead to grow with allocation sizes. 

\begin{figure}[t]
        \centering
        \includegraphics[width=0.45\textwidth]{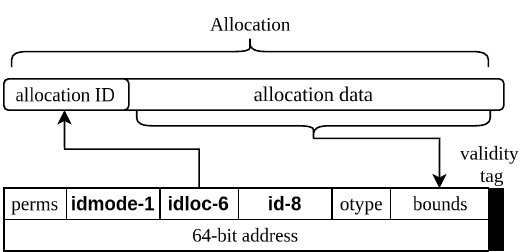}
        \caption{Inline ID mode: Memory ID placed next to the allocation}
        \label{fig:inlineid_mode}
\end{figure}
To ensure that the allocation ID can be reliably located, the ID address is not defined relative to the capability bounds, as the program may dynamically shrink these bounds ({\it e.g.} using the \emph{setBounds} instruction).
Therefore, the \emph{IDMODE} field specifies the size of an aligned region that encompasses the current bounds (and any derived bounds), and
the \emph{IDLOC} field encodes an offset within this region.

The current version of \cherid{} defines two granularities: 64-byte lines and 4-kilobyte pages.

\subsection{Inline IDs - efficiency and consistency}
\label{subsection:inline_id}
For inline IDs, the allocator places the memory ID in the final byte of the padded allocation slot. IDLOC encodes the end of this slot in 16-byte units and is calculated as

\[
\mathrm{IDLOC} =
\frac{O_{\mathrm{cache}} + S_{\mathrm{pad}}}{16},
\]

where $O_{\mathrm{cache}}$ is the allocation offset within the 64-byte aligned region and $S_{\mathrm{pad}}$ is the padded allocation size. The ID address is then obtained by

\[
A_{\mathrm{ID}} =
B_{\mathrm{reg}}
+16(\mathrm{IDLOC}\mathbin{\&}0x1f)
-1-((\mathrm{IDLOC}\gg5)\mathbin{\&}1),
\]

where $B_{\mathrm{reg}}$ is the corresponding 64-byte aligned region base. This encoding allows the ID to occupy otherwise-unused allocation padding while remaining outside the bounds of the user capability.

Inline ID storage is preferred to improve spatial locality between the inlined ID and the data.
If the ID is on the same cache line as the data being accessed, the ID can be retrieved with the data without an additional cache access. 
Indeed, inline IDs eliminate the need to buffer metadata using additional SRAM caches and tag controller support, which is necessary in previous approaches~\cite{ARM_MTE, ApplicationDataIntegrity,gulmezPICASSOScalingCHERI2026}.

Inlining the ID also provides efficient ID synchronization in multi-core systems by allowing ID updates to leverage the existing cache-coherence hierarchy. IDs are updated on deallocation to invalidate stale capabilities; because an inline ID resides in the same coherent cache line as the object, ID updates are propagated through the existing coherence mechanism.

\subsection{In-page IDs}
\label{subsection:inpage_id}

\begin{figure}[t]
        \centering
        \includegraphics[width=0.45\textwidth]{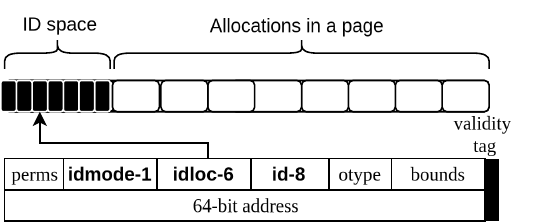}
        \caption{In-page ID mode: IDs placed within an ID table at the top of a page}
        \label{fig:inpage_id_mode}
\end{figure}

Inline ID is preferred when both the object and its ID fit within the same cache line, allowing the ID validation to accompany the normal data access. It is mostly used for objects smaller than a cache line size. When the object spans multiple cache lines or there is insufficient space for the ID in the object's cache line, \cherid{} instead uses in-page ID mode. Although this may require a separate cache-line access for the ID, placing the ID within the same physical page avoids an additional address translation.

In-page mode places IDs in a compact table at the top of each allocation page, as illustrated in Figure~\ref{fig:inpage_id_mode}. Our current design supports a maximum of 64 IDs per page, allowing the page's ID metadata to fit within a single cache line. This layout decouples ID placement from the object's cache-line layout while preserving spatial locality between metadata and data. To further reduce ID-access latency, \cherid{} caches in-page IDs in a small, virtually addressed ID buffer, described in Section~\ref{subsection:id_buffer}.

Decoding \emph{IDLOC} in in-page ID mode differs from inline mode.
Here, \emph{IDLOC} identifies an offset within the \emph{ID} table allocated at the top of a page: 
\emph{ID address = PAGE\_BASE + PAGE\_SIZE - IDLOC -1}. 
Reserving the final cache line of pages containing in-page IDs can introduce some memory overhead. The allocator can reduce this overhead by placing objects with sufficient internal fragmentation at the end of the page, allowing the ID table to occupy otherwise-unused allocation space.

\subsection{Immediate freed object reuse}
\label{sec:immediate-reuse}
\cherid{} allows allocator slots containing freed objects to be
immediately reused without waiting for a revocation sweep. When an object is freed, its memory ID is advanced before the slot is returned to the allocator. A subsequent allocation receives a capability carrying a new ID, while dangling capabilities retain the previous ID and therefore fail ID validation if dereferenced. This allows the allocator to safely reuse recently freed slots through its normal fast path, avoiding the reuse delay and memory pressure introduced by quarantine. 

Immediate reuse remains safe when the finite ID space is exhausted. With an 8-bit ID, a slot can use 254 IDs for immediate reuse; ID 255 is reserved to represent exhaustion/quarantine and is never assigned to capabilities. When the final reusable ID is freed, the memory ID is set to 255, and the slot is quarantined rather than allowing the ID to wrap. Consequently, no dangling capability can match the quarantined object. The slot becomes reusable again only after the underlying revocation mechanism has invalidated outstanding capabilities. 

Thus, \cherid{} does not eliminate sweeping revocation but moves it from the common reuse path to the ID-exhaustion case. Quarantine and sweeping remain the safe fallback, while most frees can be followed by immediate ID-based reuse.

\begin{figure}[h]
        \centering
        \includegraphics[width=0.48\textwidth]{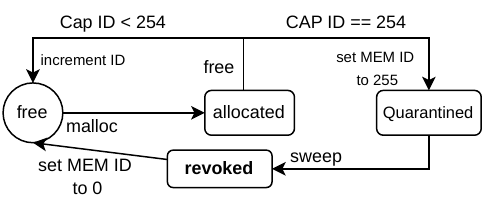}
        \caption{\cherid{} heap memory lifetime}
        \label{fig:id_flow}
\end{figure}

\subsection{Revocation without a shadow bitmap}
\label{sec:revocation-no-shadow}
Cornucopia Reloaded~\cite{Cornucopia_reloaded} uses a shadow bitmap to identify quarantined memory during revocation. The allocator paints the corresponding bitmap region on free, using one bit per capability-sized word. Although this optimizes revoker lookup, it adds allocator and memory overhead: small allocations require bit manipulation, while large allocations require painting proportionally larger regions\footnote{For 16-byte capabilities, this amounts to $allocation\_size/128$ words.}.

\cherid{} instead uses a reserved memory ID to represent quarantined allocations. This allows the revoker to determine whether a capability refers to quarantined memory directly from the allocation's ID metadata, without consulting the shadow bitmap. 
Consequently, freeing a \cherid{}-protected allocation requires only updating its ID byte, rather than painting a bitmap region proportional to the allocation size. 
Shadow-bitmap storage is also unnecessary for memory covered by \cherid{}, reducing its physical-memory overhead.

Our prototype reserves memory ID 255 to indicate quarantined memory.
When the revoker encounters a capability with a non-zero capability ID, it probes the corresponding memory ID using \emph{cgetmemID} and revokes the capability if the ID indicates quarantine.
If the capability ID is zero, the revoker falls back to the shadow bitmap for revocation using the existing Cornucopia algorithm.
This implementation integrates well with the existing load barrier mechanism of Cornucopia Reloaded, requiring only minimal changes.
The changes consist of a new predicate function that verifies whether a capability is revoked, a new support assembly routine that safely probes the memory ID in user memory, and glue logic to enable \cherid{} revocation.

Capability validation in the \cherid{} sweep requires loading IDs, which are inlined with data, while Cornucopia performs lookup purely in a shadow bitmap. 
Loading an ID from user memory can trigger a page fault, requiring the new page to be scanned before it can be mapped and accessed.
Unfortunately, the capabilities in this page may also trigger page faults, leading to problematic recursion.
In the Cornucopia case, shadow bitmap pages never hold capabilities; therefore, this additional revocation scan can be safely skipped.
These recursive scans are only a problem for pages that are paged out to swap; in other cases, it is possible to satisfy the probe as a normal non-capability read access.
Our prototype does not attempt to bound this recursion; however, as long as capabilities are not loaded from the probed page, it is possible to defer this nested scan without loss of security properties.
One simple approach would be to map the page while always trapping on capability loads, but allowing data loads; this would permit the memory ID probe while deferring the revocation scan until a later point in the main revocation memory sweep.

\section{\cherid{} Implementations}

The \cherid{} implementation effort, both in hardware and software, greatly shaped the architecture, ensuring that the \cherid{} design is eminently practical and fits conveniently into microarchitecture and allocator design.

\begin{figure}[t]
\centering
\begin{tikzpicture}[
    font=\scriptsize,
    >=latex,
    node distance=2.6mm,
    box/.style={
        draw,
        rounded corners=1pt,
        align=center,
        text width=0.36\columnwidth,
        minimum height=6mm,
        inner sep=1.5pt
    },
    small/.style={
        draw,
        rounded corners=1pt,
        align=center,
        text width=0.17\columnwidth,
        minimum height=6mm,
        inner sep=1pt
    },
    arr/.style={->, thick},
    lab/.style={font=\tiny, inner sep=0.5pt}
]


\node[font=\bfseries] (m0title) at (-2,0)
    {(a) Mode 0: Inline};

\node[box, below=2mm of m0title] (m0exe)
    {Lookup data PA at TLB}; 

\node[box, below=of m0exe] (m0l1)
    {Read cache line into L1};

\node[box, below=of m0l1] (m0check)
    {Check ID};

\node[small, below=20mm of m0check, xshift=-0.10\columnwidth] (m0fail)
    {ID fault};

\node[small, below=20mm of m0check, xshift=0.10\columnwidth] (m0failstore)
    {Silent fail};

\node[small, below=5mm of m0check, xshift=0.15\columnwidth] (m0good)
    {Proceed};








\draw[arr] (m0exe) -- (m0l1);
\draw[arr] (m0l1) -- (m0check);

\draw[arr]
    (m0check.south) -- ++(0,-17.5mm)
    -| (m0fail.north);

\draw[arr]
    (m0check.south) -- ++(0,-17.5mm)
    -| (m0failstore.north);

\draw[arr]
    (m0check.south) -- ++(0,-2.5mm)
    -| (m0good.north);

\node[above=2.8mm of m0good.north, lab]{match};
\node[below=8.75mm of m0check.south, left, lab]{mismatch};
\node[above=2.8mm of m0fail.north, lab]{load};
\node[above=2.8mm of m0failstore.north, lab]{store};






\node[font=\bfseries] (m1title) at (2,0)
    {(b) Mode 1: In-page};

\node[box, below=2mm of m1title] (m1exe)
    {Derive ID virtual address};

\node[box, below=of m1exe] (m1buf)
    {Check cached ID};

\node[box, below=of m1buf] (m1tlb)
    {Lookup data PA at TLB};

\node[box, below=of m1tlb] (m1idpa)
    {Derive ID PA from data PA};

\node[box, below=of m1idpa] (m1fallback)
    {On buffer mismatch:\\
     search LSQ $\rightarrow$ read ID from L1};

\node[box, below=of m1fallback] (m1check)
    {Compare loaded memory ID\\
     and refill ID buffer};

\node[box, below=of m1check] (m1done)
    {match: proceed\\
     mismatch: ID fault};

\draw[arr] (m1exe) -- (m1buf);

\draw[arr]
    (m1buf) --
    node[right,lab]{mismatch or not found}
    (m1tlb);

\draw[arr] (m1tlb) -- (m1idpa);
\draw[arr] (m1idpa) -- (m1fallback);
\draw[arr] (m1fallback) -- (m1check);
\draw[arr] (m1check) -- (m1done);

\draw[arr]
    (m1buf.east) -- ++(0.20,0)
    |- node[pos=.22,right,lab]{match}
    (m1done.east);

\end{tikzpicture}

\caption{
\cherid{} ID-validation paths for inline and in-page modes.
For inline loads, the ID returned with the cache line is validated in the load-response path before register writeback; for stores, the cache-line ID is validated immediately before the L1 write.
In-page mode performs a virtually addressed ID-buffer lookup before data-TLB translation. On a buffer mismatch, the object's translation is reused to derive the ID's physical address, after which pending ID updates in the LSQ and the memory-resident ID are checked.
}
\label{fig:cherid_id_paths}
\end{figure}

\subsection{\cherid{} on CHERI-Toooba}
\label{sec:hardware-implementation}
Figure~\ref{fig:cherid_id_paths} summarizes the two ID-validation paths in our CHERI-Toooba implementation. Inline mode exploits cache-line locality to validate the ID as part of the data-cache access, while in-page mode uses an ID buffer to perform validation before the data access proceeds.

\subsubsection*{Inline ID verification}
For objects protected by inline ID mode, the object data and ID reside within the same 64-byte cache line. CHERI-Toooba therefore retrieves both in a single cache access, avoiding a separate metadata lookup.

For loads, the data and ID are returned together from the cache, allowing an ID mismatch to be detected before register write back and reported as a precise exception. Stores are more challenging in out-of-order execution: delaying the commit until the ID is verified would delay retirement and strain reordering capacity, and delaying store forwarding until ID verification would constrain memory-level parallelism. Instead, our \cherid{} implementation in CHERI-Toooba silently cancels a mismatching store in the cache, not reporting an architectural exception. However, before  this store validation, a younger load may receive the value through store-to-load forwarding. Thus, although a mismatching store cannot modify memory, inline mode does not prevent consumption of its value by younger loads.

This represents a deliberate performance tradeoff in the current prototype. Software requiring stronger semantics can force allocations that would normally use inline mode to use in-page mode; our current allocator exposes this option through a software environment variable. Another possible option is to disable store-to-load forwarding only for inline-protected memory accesses, ensuring the store does not forward its value to be consumed by the load before passing an ID validation. Since this only applies to inline-protected objects, normal store-to-load forwarding remains available for in-page objects, limiting the scope of the performance tradeoff.

\subsubsection*{In-page ID verification} 
As shown in Figure~\ref{fig:cherid_id_paths}, in-page mode begins ID validation early in the memory pipeline. In \emph{MemExe}, the memory-execution stage of CHERI-Toooba, \cherid{} derives the ID virtual address from the capability's address and \emph{IDLOC}. Because the ID buffer is virtually addressed, its lookup can begin before data-TLB translation. On an ID-buffer hit, the buffered ID is compared with the capability ID; a mismatch prevents the operation from proceeding to the data access and raises a precise ID exception. 
If validation cannot be completed immediately, the operation remains pending while the ID is resolved through the fallback path. In-page mode can therefore raise precise ID exceptions for both loads and stores. Software may select in-page mode for objects that would otherwise use inline IDs when stronger store-validation semantics are required.

Since an in-page ID shares the object's physical page, its physical address can be derived from the object's translation without an additional TLB lookup; however, it requires an additional memory access to this page. \cherid{}, however, uses an ID buffer to eliminate these additional ID-accesses in most cases. 

\subsubsection*{ID buffer} 
\label{subsection:id_buffer}
To accelerate in-page ID verification, \cherid{} uses a small, virtually addressed ID buffer. 
Our implementation is a 64-entry, 4-way set-associative buffer, with each entry caching 16 object IDs. 
The buffer is indexed by the ID virtual address, derived from the object's virtual address and \emph{IDLOC}, allowing ID lookup and validation before address translation.

On a buffer miss or ID mismatch, \cherid{} checks for pending ID updates in the store queue before retrieving a word containing the ID from the cache. 
The word containing the ID is inserted into the buffer, keyed by virtual address, so that future ID validation can be done concurrently with TLB lookup.
The common case therefore requires no additional memory access,
while the fallback path requires no additional address translation.

This ID buffer mechanism is adapted from PICASSO's color-lookup implementation for CHERI-Toooba~\cite{gulmezPICASSOScalingCHERI2026}. 
Unlike PICASSO, \cherid{} co-locates metadata with its object in the same physical page, allowing the object's translation to be reused for the ID lookup, which avoids additional metadata TLB lookup, enabling a simpler memory pipeline and better common-case performance. 

Unlike PICASSO's color buffer, which flushes the entire buffer on a fence following a color update to clear stale color, \cherid{} maintains ID-buffer consistency with pending ID updates by searching the store queue during ID validation. If the buffered ID does not match the capability ID, \cherid{} first checks for a pending ID update before reloading the memory-resident ID. This synchronization with the store queue allows \cherid{} to retain valid ID-buffer entries across ordinary fences, avoiding the full-buffer flush required by PICASSO. The ID buffer is still flushed on \emph{fence.vma}, including during context switches, to prevent stale entries resulting from virtual-address reuse or aliasing.

Our prototype currently targets a single core. Multicore support requires maintaining ID-buffer coherence with ID updates from other cores. Exploring designs to efficiently integrate these entries with the physically addressed cache-coherence hierarchy remains future work.

\subsection{Allocator support for \cherid{}}
\cherid{} is a hardware-software co-design that requires allocator support  to assign and update IDs across allocation lifetimes. 
Our prototype integrates this functionality into the Malloc Revocation Shim (MRS)~\cite{Cornucopia_reloaded} over CheriBSD jemalloc. This integration allows us to evaluate \cherid{} within a realistic CHERI software stack, including its effects on memory reuse, revocation behavior, and system performance.
Implementing ID management in MRS adds prototype instruction overhead; tighter integration with jemalloc could reduce this cost.

\subsubsection*{Support for inline ID}
The \cherid{} inline ID mode co-locates the memory ID with allocation data in the same cache line. The ID is placed immediately above the user-requested object but outside its capability bounds, protecting the ID from modification by the application.

As shown in Table~\ref{table:allocation_distribution}, a high proportion of the allocation requests running on CheriBSD, when serviced by jemalloc, exhibit internal fragmentation due to the rounding of the allocator size-class.
Our prototype opportunistically places inline IDs within this otherwise-unused space, avoiding additional memory overhead. When the requested size exactly matches a size class and therefore provides no space for an ID, the allocator selects the next larger size class.

\subsubsection*{Support for in-page ID}
The \cherid{} in-page ID mode requires the allocator to reserve space for an ID table at the top of each allocation page.
Our prototype reserves 64 bytes, providing one byte for each of the 64 ID locations addressable by IDLOC.

For implementation simplicity, we modify jemalloc's allocation logic to mark allocation slots that overlap the reserved ID region as unavailable. This prevents the allocator from placing user objects over ID metadata, but may leave the remainder of affected slots unused and therefore introduces fragmentation. A more space-efficient implementation could instead place objects with sufficient internal fragmentation adjacent to the ID table, allowing the metadata to occupy otherwise-unused padding.

\section{Evaluation}
We evaluate \cherid{} along three dimensions: temporal-safety enforcement, performance and memory-system behaviour, and hardware cost.

\begin{itemize}
    \item To evaluate \cherid{}'s temporal safety enforcement, we use version 1.3 of the U.S. NIST SARD Juliet Test Suite~\cite{Juliet}.
     \item To evaluate application-level performance and memory-system behavior, we conduct FPGA-based experiments using SPEC CPU2006 INT, gRPC, and SQLite workloads. 
     We further use targeted allocator and revocation microbenchmarks to characterize allocation/free costs and the cache and TLB pressure introduced by quarantine and revocation.
     \item To evaluate hardware cost, we report FPGA resource utilization of the \cherid{} extensions relative to the baseline CHERI-Toooba.
   
\end{itemize}

\subsection{Experimental setup}

\cherid{} is a hardware-software co-design spanning the CHERI software and hardware stacks. Our prototype extends the LLVM compiler toolchain, CheriBSD, and the jemalloc user-space allocator, together with the QEMU emulator and CHERI-Toooba processor implementation.

For performance and hardware evaluations, we extend CHERI-Toooba~\cite{ruggSuiteProcessorsExplore2024} with \cherid{} and deploy the modified system on a Xilinx VCU118 FPGA. The system is configured with a 4-way, 32 KB L1 data cache, and a 16-way, 1 MB last-level cache connected to the CHERI tag controller and DRAM.
We synthesized our design using Vivado 2019 and run the prototype at 25 MHz. We use the processor's hardware performance counters to measure execution cycles and memory-system behavior, including cache and TLB activity.
We additionally implement \cherid{} in QEMU to support rapid hardware--software co-design and functional validation.

\subsection{Hardware Overheads}
\label{sec:hardware-overheads}
\begin{table}[t]
\centering
\caption{FPGA resource utilization and overheads relative to baseline.}
\label{tab:fpga-area}
\begin{tabular}{lrrr}
\toprule
Resource & Baseline & \cherid{} & PoisonCap \\
\midrule
Logic LUTs  & 687,633 & 2.10\% & 1.10\% \\
Registers   & 419,043 & 2.64\% & 2.20\% \\
Memory LUTs & 20,075  & 7.66\% & 0.10\% \\
\bottomrule
\end{tabular}
\end{table}

We evaluate FPGA resource utilization as a proxy for the logic and memory complexity of the \cherid{} implementation. We additionally compare against PoisonCap~\cite{wangPoisonCapEfficientHierarchical2026}, a recent CHERI temporal-safety design that poisons freed memory and validates the poison state on subsequent memory accesses.
\cherid{}'s ID tracking and validation mechanisms incur modest additional hardware costs.
Relative to baseline CHERI-Toooba, as shown in Table~\ref{tab:fpga-area}, \cherid{} increases combinational logic (LUT) usage by +2.1\%, flip-flop logic (registers) by +2.6\%, and memory resource utilization by +7.7\%, compared with PoisonCap's +1.1\%, +2.2\%, and +0.1\%, respectively. PoisonCap is therefore more resource-efficient, particularly in memory LUTs; \cherid{} caches separately stored IDs in an ID buffer, but PoisonCap validates poisoned state in place.

\subsection{Quarantine and Revocation Overheads}
\label{sec:quarantine-revocation-overhead}
To quantify the memory-system costs of quarantine, we construct a
single-threaded allocator microbenchmark that repeatedly allocates and frees 31-B objects for 100,000 iterations on FPGA, while collecting Toooba hardware performance counters. This workload exercises steady-state allocation/reuse without triggering revocation, 
allowing us to isolate the cost of quarantine. Cornucopia incurs 2.91$\times$ the baseline cycles, 19.4$\times$ DTLB misses, 19.7$\times$ L1 D-cache misses, and 18.3$\times$ last-level cache (LLC) misses. 
In comparison, \cherid{} incurs only 1.46$\times$, 1.91$\times$, 1.65$\times$, and 1.36$\times$, respectively. These results demonstrate that quarantine alone substantially increases cache and TLB pressure even without revocation, and \cherid{} can effectively reduce quarantine overhead. 

We separately measure revocation sweep overhead using a controlled microbenchmark
that first allocates and frees 100,000 64-B objects to populate the quarantine, and then explicitly calls \texttt{malloc\_revoke()} to capture the incremental cost of the revocation sweep.
The sweep increases execution cycles by 1.39$\times$, instructions by 1.42$\times$, DTLB accesses by 1.35$\times$, and LLC misses by 1.62$\times$, while LLC-miss wait cycles increase by 1.70$\times$. These results show that revocation scanning introduces substantial CPU and memory-system costs beyond quarantine itself. \cherid{} reduces these costs by eliminating shadow-bitmap painting for protected allocations, enabling immediate memory reuse, and substantially reducing sweep frequency.

\subsection{Allocator performance}
\label{sec:allocator-performance}
We evaluate allocator performance using a microbenchmark that generates repeated \texttt{malloc}/\texttt{free} operations running on FPGA, covering both inline (15/31 B) and in-page (64 B--2 KiB) ID modes, as shown in Table~\ref{tab:alloc-micro}. \cherid{} reduces allocation/reuse costs by 1.9--4.3$\times$ relative to Cornucopia. Notably, the transition from inline to in-page IDs introduces little performance discontinuity.
\begin{table}[t]
\centering
\caption{Allocation/free latency (cycles per pair) on FPGA.}
\label{tab:alloc-micro}
\begin{tabular}{lrrrrr}
\toprule
& \multicolumn{2}{c}{Inline ID} &
  \multicolumn{3}{c}{In-page ID} \\
\cmidrule(lr){2-3}\cmidrule(lr){4-6}
& 15 B & 31 B & 64 B & 512 B & 2 KiB \\
\midrule
Baseline(no temporal)    & 199 & 210 & 200 & 209 & 186 \\
Cornucopia  & 584 & 612 & 652 & 967 & 1746 \\
\cherid{}     & 301 & 308 & 315 & 310 & 403 \\
\bottomrule
\end{tabular}
\end{table}

\begin{figure*}[t]
\centering
\includegraphics[width=0.48\textwidth]{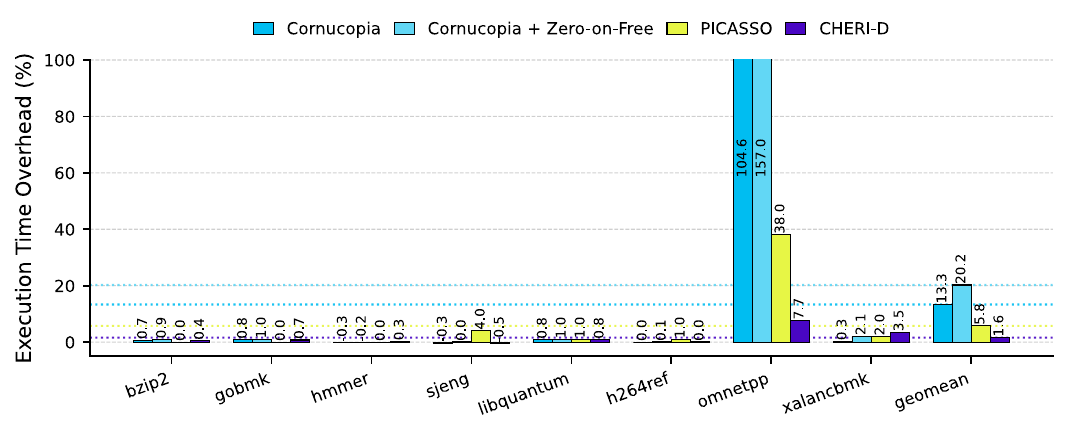}
\hfill
\includegraphics[width=0.48\textwidth]{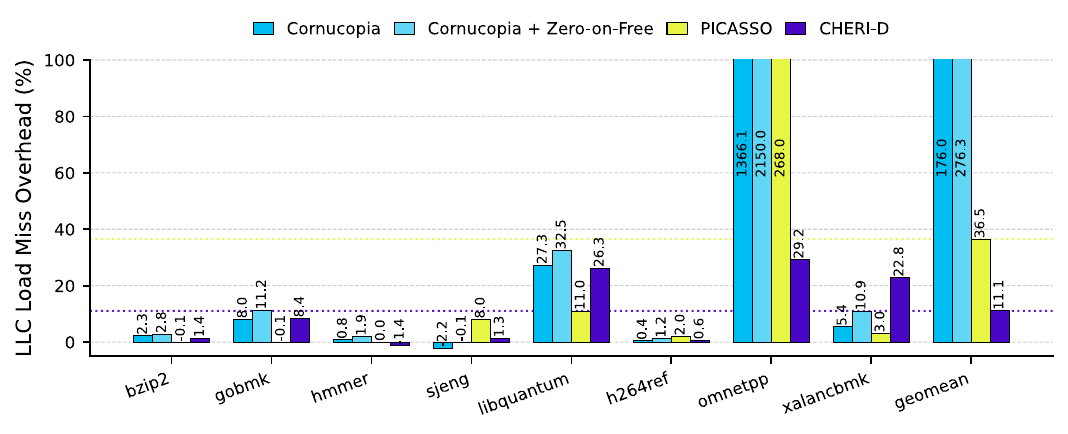}
\caption{Execution time overhead (left) and DRAM traffic (last-level cache miss) overhead (right) of SPEC CPU2006 integer benchmarks on the \cherid{} FPGA-based prototype. The DRAM traffic results are shown for a configuration in which \cherid{} supports allocations smaller than 64 bytes.}
\label{fig:fpga_spec_overheads}
\end{figure*}

\subsection{Security}
For the security evaluation, we use version 1.3 of the U.S.
NIST SARD Juliet Test Suite~\cite{Juliet,Juliet_test_doc}, which contains thousands of C/C++ related memory vulnerabilities.
To verify that \cherid{} can mitigate temporal safety violations, we tested the following relevant classes: CWE-415: Double Free and CWE-416: Use After Free from the Juliet Test Suites, which have a total of 2422 tests, against our prototype implementations on both FPGA and QEMU.

\subsubsection*{Use-After-Free}
Both our FPGA-based and QEMU-based \cherid{} prototype implementations have successfully executed all provided CWE-416 good test cases, and detected and trapped all 416 vulnerabilities in the ``bad'' cases. 
These results demonstrate that the architectural mechanism introduced by \cherid{} enables CHERI to provide deterministic detection of the evaluated heap use-after-free accesses, extending beyond Cornucopia Reloaded's use-after-reallocation mitigation~\cite{wangPoisonCapEfficientHierarchical2026}.

\subsubsection*{Double Free}
Our \cherid{} prototype implementations have also successfully passed all 1636 tests in the CWE-415: Double Free class of the Juliet Test Suite~\cite{Juliet}.
\cherid{} does not rely on the shadow bitmap to detect double-free, which is essential for Cornucopia Reloaded. \cherid{} detects heap double-free by checking if the allocation ID has been incremented on free by comparing the allocation ID against the ID in the capability being freed. If the memory ID no longer matches the capability ID being freed, the capability refers to an earlier allocation lifetime, indicating a repeated or stale free.

\begin{figure*}[t]
   \centering
    \includegraphics[width=\textwidth]{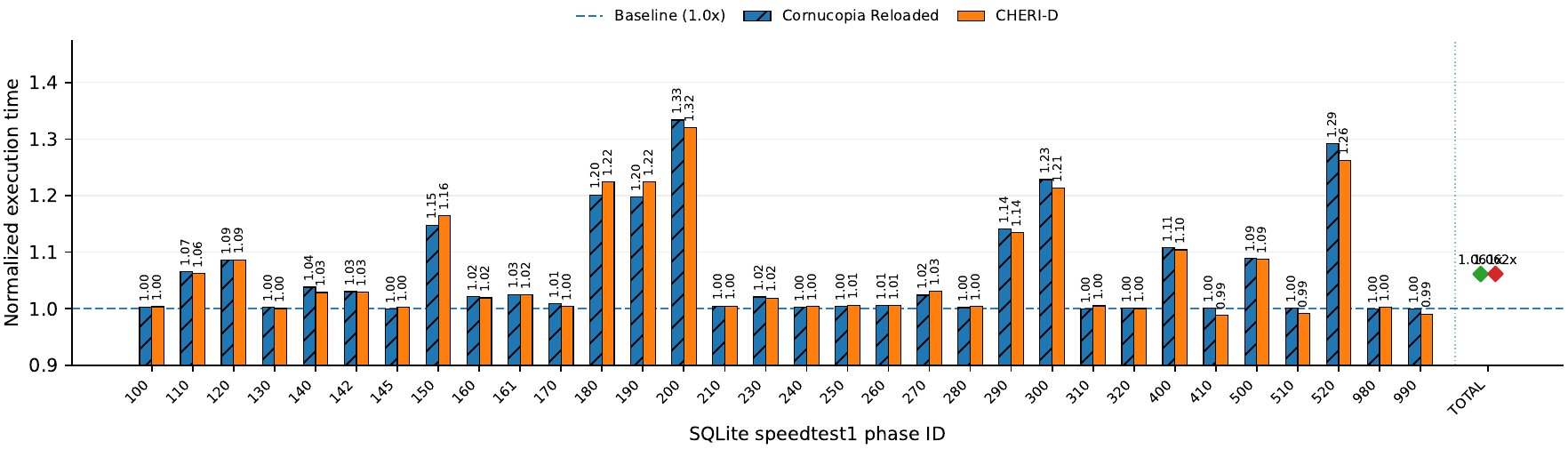}
     \caption{Normalized Performance overhead of Cornucopia Reloaded (1.07x avg., 1.33x max.) and \cherid{} (1.06x avg., 1.32x max.) across SQLite speedtest1 operational phases.}
    \label{fig:sqlite_perf}
\end{figure*}

\begin{figure}[t]
        \centering
        \includegraphics[width=0.5\textwidth]{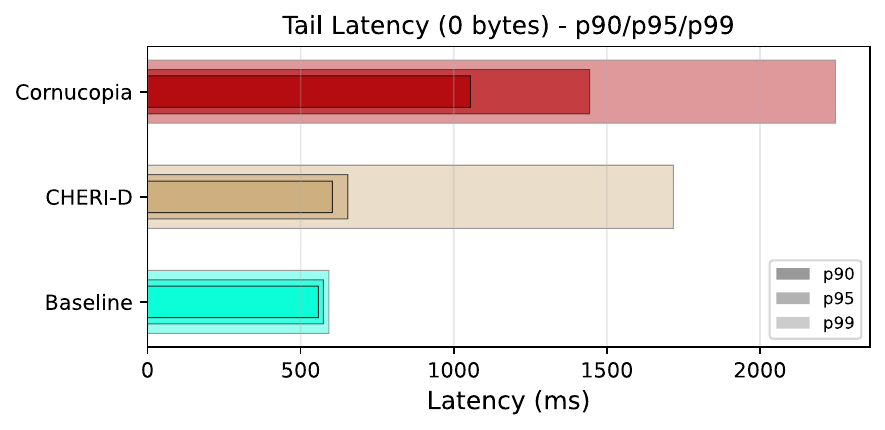}
        \caption{ gRPC QPS latency percentile.}
        \label{fig:gRPC_latency}
\end{figure}

\subsection{Performance Evaluation on FPGA}
\label{sec:performance-evaluation}
\subsubsection*{SPEC CPU2006 INT}
To evaluate the performance of \cherid{}, we used the CHERI-supported subset of the SPEC CPU2006 INT benchmarks~\cite{henningSPECCPU2006Benchmark2006}.
Like earlier performance evaluations on CHERI-Toooba~\cite{ruggEfficientSpatialTemporal}, we primarily used the \emph{train} input configuration for FPGA-based experiments.
The inline-ID mode is enabled for allocations under a cache line size (64 bytes), and the in-page ID mode is enabled for allocations under a page size (4096 bytes).
To expose the full revocation overhead, we did not enable concurrent revocation; both the application and the revoker execute on the same hardware thread throughout the evaluations. 
Figure~\ref{fig:fpga_spec_overheads} shows that workloads that are not allocation sensitive, and therefore trigger revocation infrequently, incur only a moderate performance overhead on Cornucopia Reloaded.
However, workloads characterized by frequent memory allocation and deallocation operations experience substantially higher overhead due to the combined costs of memory quarantining and sweeping.
Among the evaluated benchmarks, \textit{omnetpp} exhibits the highest sensitivity to these mechanisms.
As shown in Table~\ref{table:allocation_distribution}, \textit{omnetpp} issues approximately 130 million memory allocation requests, and a similar number for deallocation. 
Consequently, as reported in Table~\ref{tab:revocation-quarantine}, a total of {2792} revocation sweeps are triggered under Cornucopia Reloaded. 
As reflected in Figure~\ref{fig:fpga_spec_overheads}, \textit{omnetpp} experiences a runtime overhead of {$104.6\%$}.
Performance degradation is dominated by memory-system interference, as evidenced by $13\times$ increases in LLC misses or DRAM traffic.  

By enabling immediate memory reuse, \cherid{} substantially mitigates these costs. 
Specifically, the number of revocation sweeps triggered during \textit{omnetpp} execution is dramatically reduced, from {2792} to just {12}. 
PICASSO similarly seeks to reduce the frequency of the revocation sweep through immediate object reuse~\cite{gulmezPICASSOScalingCHERI2026}.
Unlike \cherid{}, PICASSO is limited in the number of live allocations it can support; this is due to the maximum width of the color encoded in the capability, which determines the size of the color table and puts an upper bound of approximately 2.09 million live allocations.
Once the available colors are exhausted, a revocation sweep must be performed to recover color identifiers.
This fixed color space limits the maximum number of simultaneously protected allocations in the evaluated PICASSO configuration.
Under \textit{omnetpp} with the \emph{train} input set, PICASSO triggers {70} revocation sweeps and incurs higher performance overhead and DRAM traffic than \cherid{}, as shown in Figure~\ref{fig:fpga_spec_overheads}. 
Furthermore, PICASSO is unable to execute \textit{omnetpp} with the \emph{ref} input set because the benchmark reaches a peak of approximately 2.17 million maximum live allocations, exceeding PICASSO's maximum supported allocation count.

Despite these advantages, both PICASSO and \cherid{} exhibit a higher performance overhead than Cornucopia Reloaded when executing \textit{xalancbmk}.  
Although {$96.7\%$} of the allocation requests are for objects smaller than {4096 bytes}, {$81.01\%$} of the total heap memory consumption is contributed by {$3.06\%$} of the allocation requests that are larger than {4096} bytes.
These large allocation requests exceed the maximum object size currently supported by our implementation of \cherid{}, preventing the design from fully exploiting immediate memory reuse and consequently limiting its performance benefits for \textit{xalancbmk}.

In general, as shown in Figure~\ref{fig:fpga_spec_overheads}, \cherid{} incurs only a {$1.6\%$} average performance overhead compared to {$13.3\%$} and {$5.8\%$} for Cornucopia Reloaded and PICASSO, respectively.
Similarly, \cherid{} increases DRAM traffic by only {$11.1\%$} on average, while Cornucopia Reloaded and PICASSO increase DRAM traffic by {$176\%$} and {$36.5\%$}, respectively. These results demonstrate that \cherid{} can substantially reduce both execution-time and memory-system overhead by enabling immediate memory reuse and significantly reducing the frequency of revocation. 
Furthermore, although the current FPGA-based \cherid{} prototype supports only allocations smaller than a page, it nevertheless scales effectively to allocation-intensive workloads such as \textit{omnetpp}. 

As PoisonCap was evaluated using a dual-core
configuration, we could not directly compare its reported performance
with our single-core results. As an approximate reference, we evaluate Cornucopia Reloaded with zeroing on free, which PoisonCap reports to having similar performance~\cite{wangPoisonCapEfficientHierarchical2026}. As shown in Figure~\ref{fig:fpga_spec_overheads}, this configuration incurs an average higher performance overhead than \cherid{}.

\subsubsection*{SQLite}
We evaluated \cherid{} using SQLite's built-in speedtest1 benchmark, which comprises multiple phases of database operations such as create, insert, reorder, and delete. As shown in Table~\ref{table:allocation_distribution}, the benchmark exhibits a high proportion ({$65.8\%$}) of requests for large heap allocations (approximately 4 KiB – 4 MiB), which contributes to {$99.53\%$} of the total heap size. This distribution arises because SQLite's lookaside allocator services most small allocation requests~\cite{DynamicMemoryAllocation}, leaving the general-purpose heap dominated by larger objects. Since these allocation request sizes exceed the maximum allocation size currently supported by \cherid{}, it provides only limited optimization for this workload, benefiting only {$0.47\%$} of heap memory. Despite this limitation, \cherid{} reduced the total number of revocation sweeps from {267} to {242}, which improves performance for 19 of the 32 speedtest1 tests, as shown in Figure~\ref{fig:sqlite_perf}. 
PICASSO, in contrast, supports larger allocations and eliminates sweeps for this workload but still incurs approximately $8\%$ average performance overhead. PICASSO requires an additional metadata-address translation and lookup on every color miss, including small objects. In contrast, \cherid{} co-locates ID metadata with allocation data and shares the object's existing address translation. This difference introduces a common-case tradeoff: PICASSO achieves broader allocation coverage and eliminates revocation but introduces common-case memory access overhead, whereas \cherid{} retains lower metadata access costs and consequently achieves lower average overhead despite the remaining revocation sweeps.
These results motivate extending \cherid{} to larger allocations.
Large allocations account for $99.53\%$ of SQLite's allocated heap
memory and contribute substantially to its remaining revocation
pressure. Supporting them with a more flexible ID encoding could
further reduce revocation while retaining \cherid{}'s common-case
metadata-access efficiency.

We additionally measure the quarantine memory overhead for both SPEC CPU2006 and SQLite, as shown in Table~\ref{tab:revocation-quarantine}.  Although \cherid{} substantially reduces quarantine frequency and average overhead, it does not reduce the maximum quarantine size under a fixed revocation threshold. After exhausting 254 reusable IDs, an allocation slot enters quarantine until the threshold triggers a revocation sweep. Thus, \cherid{} reduces how frequently and how long memory is quarantined, rather than the maximum quarantine size.

\subsubsection*{gRPC QPS}
We evaluated gRPC 1.54.2 using the QPS benchmark framework~\cite{GrpcGrpcV1542}.
The gRPC server executes on CHERI-Toooba on FPGA, while benchmark clients run on a host system to isolate the server-side allocator and revocation overheads from client-side execution effects, allowing us to measure latency percentiles. 
We use a custom scenario to minimize application-level payload processing and reduce execution variability by fixing request/response sizes to 0 bytes, disabling transport security, and using a single synchronous client-server configuration with one channel and eight unary RPCs in a closed-loop execution model. Each experiment consists of a 15s warm-up period, followed by a 120s measurement phase.

We evaluated gRPC with three CHERI temporal safety configurations: baseline with revocation disabled, Cornucopia Reloaded, and \cherid{}. 
Figure~\ref{fig:gRPC_latency} presents the results of the latency percentile. At the p90 latency percentile, \cherid{} incurs an overhead of {$8.3\%$}, while Cornucopia Reloaded incurs an overhead of {$89\%$}. At the p95 latency percentile, \cherid{} incurs a {$13.8\%$} overhead, compared to {$151.3\%$} for Cornucopia Reloaded. At the p99 latency percentile, \cherid{} incurs a {$190\%$} overhead, while Cornucopia Reloaded experiences a {$279.4\%$} overhead.
The remaining p99 overhead is primarily due to allocations larger than a page, which are not covered by the current \cherid{} prototype and therefore remain subject to stop-the-world revocation on the single-threaded critical path.
 
In contrast, PICASSO supports these larger allocations and  triggers no revocation sweeps during this workload, resulting in only approximately $3\%$ QPS degradation. This demonstrates that extending \cherid{} to larger allocations is important for latency-sensitive workloads, where eliminating stop-the-world revocation can outweigh additional common-case metadata-access costs.

\begin{table}[t]
\centering
\caption{Revocation and quarantine behavior of Cornucopia and \cherid{}.
Q-Count is the number of physical quarantine events, and Q-Ratio is the time-weighted average quarantine ratio $Q(t)/A(t)$. SPEC CPU2006 results are train/ref.}
\label{tab:revocation-quarantine}

\setlength{\tabcolsep}{2.5pt}
\renewcommand{\arraystretch}{1.05}
\scriptsize

\begin{tabular}{l rrr rrr}
\toprule
& \multicolumn{3}{c}{\textbf{Cornucopia}}
& \multicolumn{3}{c}{\textbf{\cherid{}}} \\
\cmidrule(lr){2-4}
\cmidrule(lr){5-7}

\textbf{Benchmark}
& \textbf{Revoke} & \textbf{Q-Count} & \textbf{Q-Ratio}
& \textbf{Revoke} & \textbf{Q-Count} & \textbf{Q-Ratio} \\
\midrule

\multicolumn{7}{l}{\textit{SPEC CPU2006 (train/ref)}} \\

bzip2
& 2/0 & 24/24 & 22.1/0.9
& 2/0 & 24/24 & 21.8/0.9 \\

gobmk
& 11/23 & 51K/133K & 39.2/41.1
& 6/2 & 28.3K/82.5K & 39.1/40.1 \\

hmmer
& 16/124 & 170K/1.0M & 67.4/59.4
& 1/2 & 693/3.9K & 71.3/63.4 \\

sjeng
& 0/5 & 1/1 & 0/0
& 0/0 & 1/1 & 0/0 \\

libquantum
& 1/11 & 107/149 & 37.4/36.3
& 1/11 & 58/57 & 37.1/23.9 \\

h264ref
& 12/12 & 38K/38K & 38.3/38.4
& 12/12 & 5.1K/5.1K & 37.3/37.4 \\

omnetpp
& 2792/608 & 130M/267M & 40.2/39.4
& 12/6 & 151K/304K & 28.5/32.3 \\

astar
& 38/187 & 347K/1.1M & 56.0/51.2
& 9/27 & 2.7K/12.9K & 55.2/51.6 \\

xalancbmk
& 2/291 & 1.1M/135M & 8.4/38.0
& 0/9 & 33.7K/2.45M & 0.7/41.6 \\

\midrule
\multicolumn{7}{l}{\textit{SQLite}} \\

speedtest1
& 267 & 221K & 64.4
& 242 & 146K & 64.2 \\

\bottomrule
\end{tabular}
\end{table}

\section{Related work}
\label{sec:related-work}
\subsubsection*{Memory tagging}
Arm MTE associates a 4-bit tag with each 16-byte memory granule and a corresponding tag with the pointer~\cite{ARM_MTE}. Its small tag space makes tag reuse unavoidable, providing probabilistic rather than deterministic temporal protection. MTE supports synchronous and asynchronous checking, trading precise fault reporting for performance, and incurring additional costs for tag storage and manipulation.
An important advantage of MTE is compatibility with conventional 64-bit AArch64 binaries, stashing tags in unused upper pointer bits.
\cherid{} is similarly compatible with CHERI-RISC-V binaries, not requiring a change in Application Binary Interface (ABI).
However, while MTE requires the allocator to ``paint" the ID shadow space, \cherid{} requires the allocator to adjust heap layout to accommodate IDs.
In exchange, \cherid{} requires only one ID per allocation rather than metadata for every 16-byte granule, and supports an 8-bit ID, permitting 254 immediate-reuse generations before quarantine.

\subsubsection*{Inline object metadata}
PTAuth~\cite{farkhani2021ptauth} stores a 64-bit random ID before each heap object and authenticates pointers using Arm PAC, reporting 26\% SPEC CPU2006 overhead (20\% estimated with hardware PAC) and 2\% memory overhead.
\cherid{} improves on PTAuth by protecting the ID in memory using CHERI bounds, which also allows safe storage in allocator internal fragmentation, reducing memory overhead.
This deterministic protection also allows \cherid{} to use only 8 bits for IDs, as it does not rely on cryptography, further reducing memory overhead.
The simplicity afforded by relying on CHERI primitives also allows \cherid{} to implement all checking in hardware, eliminating the need for special code generation and greatly improving performance.

\subsubsection*{CHERI revocation}
CHERIvoke introduced sweeping capability revocation for CHERI, with Cornucopia, Cornucopia Reloaded, and CHERIoT subsequently developing related quarantine-and-revocation mechanisms~\cite{CHERIvoke, wesleyfilardoCornucopiaTemporalSafety2020, Cornucopia_reloaded}.
These approaches generally prevent the immediate reuse of freed memory by quarantining it until dangling capabilities can be identified and revoked. 
\cherid{} is complementary to this revocation model rather than replacing it, which reduces both quarantine pressure and the frequency with which expensive revocation sweeps are required.

\subsubsection*{Hardware table-based solutions}
PICASSO provides strict temporal safety using a per-address-space object validity table consulted on heap accesses~\cite{gulmezPICASSOScalingCHERI2026}.
Its centralized table requires additional metadata address translation and synchronization, and supports approximately two million objects.
In contrast, \cherid{} co-locates IDs with allocation data, avoiding a separate ID translation and allowing metadata storage to scale with the heap. 
Our FPGA prototype reuses PICASSO's lookup mechanism to implement in-page ID checks.

\subsubsection*{Memory poisoning}
PoisonCap extends Cornucopia Reloaded with hierarchical use-after-free mitigation and initialization safety~\cite{wangPoisonCapEfficientHierarchical2026}.
Like \cherid{}, it stores metadata within the allocation but retains quarantine rather than enabling immediate reuse, resulting in limited performance improvement over Cornucopia Reloaded. \cherid{} instead avoids quarantining in the common-case and reduces revocation frequency. 
PoisonCap also provides a broader security model covering multiple allocation layers and initialization safety.

\section{Future work}
\label{sec:future-work}
\subsubsection*{Coherent ID buffer}
Our current prototype is single-core. Local ID-buffer consistency is maintained by checking pending ID stores in the load/store queue before validation; ordinary fences, therefore do not require an ID-buffer flush. Multicore support additionally requires stale ID-buffer entries to be invalidated when another core updates an ID. 

The primary challenge to buffer coherence is due to the ID buffer being virtually addressed, whereas cache coherence operates on physical addresses.
As a result, it is unclear how remote ID updates can be associated with stale ID-buffer entries. 
If a multicore design could associate entries with physical-address information for coherence-driven invalidation, existing coherence invalidations could be reused. We leave implementation and evaluation to future work.

\subsubsection*{Larger allocation support}
Extending \cherid{} to efficiently cover allocations larger than a 4 KiB page size is an important direction for future work.
The current fixed division of capability bits between ID and IDLOC limits the allocation range covered by the in-page ID modes. 
Supporting larger allocations will require a more flexible encoding that can dynamically trade ID bits for IDLOC bits according to the allocation size. 
We plan to explore such flexible encoding to extend \cherid{} to larger allocations and further reduce
revocation pressure.

\subsubsection*{Hierarchical temporal safety support}
PoisonCap~\cite{wangPoisonCapEfficientHierarchical2026} provides a hierarchical and scalable mechanism to enforce use-after-free (UAF) mitigation across nested allocation layers. \cherid{} allows for more flexible ID metadata placement and has the potential to enable nestable schemes to support sub-allocations; however, its interaction with multiple levels of nested allocator hierarchies has not yet been fully explored. Designing effective mechanisms for preserving temporal safety across allocation layers remains an important direction for future work.

\section{Conclusion}
\cherid{} provides a foundation for practical and efficient CHERI temporal safety by extending CHERI architectural protection to allocation lifetimes.
More broadly, \cherid{} demonstrates how CHERI's spatial protection enables trusted temporal metadata to be safely co-located with application objects, relaxing the conventional requirement for a separate shadow metadata structure.
By enabling immediate memory reuse, \cherid{} reduces quarantine and revocation overhead while providing stronger temporal safety guarantees.
Our evaluation demonstrates the performance, security, and hardware feasibility of this approach, highlighting \cherid{} as a promising foundation for future CHERI temporal-safety architectures.


 use section* for acknowledgment
\ifCLASSOPTIONcompsoc
\section*{Acknowledgments}
This work was sponsored by UKRI3001: CHERI Research Centre.
Distribution Statement A: Approved for public release; distribution is unlimited. This work is also sponsored in part by the Defense Advanced Research Projects Agency (DARPA) under contract FA8750-24-C-B047 (“DEC”) as part of the CPM program. The views, opinions, and/or findings contained in this report are those of the authors and should not be interpreted as representing the official views or policies, either expressed or implied, of the Department of Defense or the U.S. Government.
We thank our colleagues on the CHERI team from the University of Cambridge for building the CHERI research platforms, and especially Jessica Clark for her assistance in getting our experiments up and running. We are also grateful to Merve Gulmez and Thomas Nyman at Ericsson Security Research for their help in reproducing the evaluations they had previously conducted.
\else
\fi


\appendices



%
\bibliographystyle{IEEEtran}
\bibliography{MyLibrary}

\newpage 

\end{document}